\documentstyle[12pt]{article}

\def\be{\begin{equation}}
\def\bea{\begin{eqnarray}}

\def\ee{\end{equation}}
\def\eea{\end{eqnarray}}
\def\R{\rm {I\kern-.200em R}}
\def\C{\rm {I\kern-.520em C}}
\begin{document}

\begin{titlepage}
\begin {center}

{\large { Quantum group symmetry of the Quantum Hall effect \\ on the non-flat
surfaces }}
\\
\vskip 1cm{M. Alimohammadi $^{1,2}$ and A.Shafei Deh Abad $^{1,3}$ }\\
\vskip 2cm
 {\it $^1$ Institute for Studies in Theoretical Physics and
Mathematics\\  P.O.Box 19395-5746 , Tehran , Iran\\
 $^2$ Physics Department , University of Tehran , North Karegar ,\\
 Tehran , Iran\\
 $^3$ Department of Mathematics , University of Tehran , \\ Tehran , Iran \\
}
\end{center}
\vskip 1cm
\begin{abstract}
After showing that the magnetic translation operators are not the symmetries of
the QHE on non-flat surfaces , we show that there exist another set of
operators
which leads to the quantum group symmetries for some of these surfaces . As a
first example we show that the $su(2)$ symmetry of the QHE on sphere leads to
$su_q(2)$ algebra in the equator . We explain this result by a contraction
of $su(2)$ . Secondly , with the help of the symmetry operators of QHE on the
Pioncare upper half plane , we will show that the ground state wave functions
form a representation of the $su_q(2)$ algebra .
\end{abstract}
\vskip 2cm
\vfill
\end{titlepage}
\noindent

\section{Introduction}

After the discovery of the quantum Hall effect (QHE) $^{[1]}$ and the
fractional
quantum Hall effect (FQHE) $^{[2]}$ , Laughlin introduced his interacting
electrons model and showed that the incompressible quantum fluid can explain
the
appearance
of the plateaus in FQHE in filling factor $\nu = {1\over m}$, m an odd integer
$^{[3]}$ . In recent
years there was many efforts to explain this feature of incompressibility by
the symmetries of the quantum mechanics of the two dimensional planar motion of
a nonrelativistic particle in a uniform magnetic field. Recently I.I.Kogan
$^{[4]}$
and H.T.Sato $^{[5]}$ , by using the magnetic translation operator showed
that there exists a
quantum group symmetry in this problem. They found that
the following combination of the magnetic translation operator
,  $T_ {\bf{a}}= exp ({\bf{a}}.({\vec{\nabla}} + i{\bf{A}}))$, where ${\bf{a}}$
is a constant vector and ${\bf{A}}$ is electromagnetic potential, could
represent the $su_q(2)$ algebra:

$$J_\pm = {1\over {q-q^{-1}}} (\alpha _{\pm} T_{\pm {\bf{a}}} + \beta _{\pm} T_
{\pm {\bf{b}}}) \ \ \, \ \ \ \hskip 1cm q^{2J_3}= T_{{\bf{b}} -{\bf{a}}}$$
where $q=exp (i{{\bf{B}}\over 2}. ({\bf{a}}\times {\bf{b}}))$ and $\alpha _+
\beta _- = \beta _+ \alpha _- = -1$ . Let $W _{n,\bar{n}}=
T_{{\bf{a}}}$
where $a_i= \epsilon _{ij} n_j$ . It can be shown that $W_{n,\bar{n}}$
satisfies
the Fairlie-Fletcher-Zachos (FFZ) trigonometric algebra $^{[6]}$ . This algebra
in the weak
field limit $(B \rightarrow 0)$ leads to $w_\infty$ algebra, the algebra of the
area-preserving diffeomorphism $^{[7]}$ . Therefore this $su_q(2)$ symmetry
indicates the incompressibility feature of the FQHE . The same symmetry
was also found for
the topological torus $^{[8]}$ .

In this paper we will study the same quantum group symmetry
for non flat surfaces. In section 2 we will show that the magnetic traslation
operators are the symmetries of the Hamiltonian only when the metric is flat .
Therefore in the case of non flat surfaces we first must look for the  symmetry
operators and after that try to find any quantum group stucture of it . In
section 3 we will begin with the first non-trivial surface , the sphere .
 The problem of the motion of the electrons in the presence
of the magnetic monopole and when the electrons are restricted to move on a
sphere , was
first considered by Haldane $^{[9]}$ . He formulated the problem such that the
symmetry algebra
of the Hamiltonian is $su(2)$ algebra , with the generators which are
represented
by a special combination of the rotation and gauge transformation . We will
consider
the group elements of this algebra with non contant parameters . That is
the set of maps from $S^2$ to $SU(2)$ . By studying its multiplication law
we will recover the FFZ algebra for a
special region of the sphere . We will try to explain this appearance
of $su_q(2)$ from $su(2)$ by a special contraction of $su(2)$ .

It is well known that the group of automorphisms of any genus $g\geq 2$ compact
Riemann surface is discrete . So to look for any quantum group symmetry we
have to consider non-compact surfaces . For this purpose we consider
the Poincare upper half plane
in section 3 , and we will show that the $su_q(2)$ algebra is the
symmetry of this surface.

After the writing of this paper was nearly finished , we was aware about the
recent preprint [10] in
which the quantum group symmetry on the sphere had been discussed .

\section {Symmetry properties of the magnetic translation operator}

Consider a particle on a Riemann surface interacting with a monopole field .
That is the integral of the field strength out of the surface is different from
zero . The natural definition of the constant magnetic field is $^{[11]}$
 $$F_{\mu \nu}=B\sqrt {g} \epsilon _{\mu \nu} \ \ ,$$
and the Hamiltonian of the electron is given by
\be H= {1\over {2m}}{1\over {\sqrt g}}(\partial _\mu -iA_\mu) \sqrt g
g ^{\mu \nu}(\partial _\nu -iA _\nu)
= {1\over {2m}} {\vec{\nabla}}^2 + {B\over {2m}}  \ \ \ , \ee
where $\nabla _\mu=\partial _\mu -iA_\mu$ . In Ref. (11)
this Hamiltonian was solved by choosing some special
metrics .
Now consider the magnetic translation operator
$T_{\vec{\xi}}=e^{{\xi^\mu}D_\mu}
$ which acts on scalars . Here ${\vec{\xi}}=\xi ^{\mu} \partial _ {\mu}$ is a
vector field
and $ D_\mu = \partial _\mu +i A_\mu$ .  The operators $T_{\vec{\xi}}$ is a
symmetry operator only when
\be [D_\mu , \nabla_\nu]=-\partial_\mu A_\nu - \partial_\nu A_\mu=0\ee
In other words we must be able to choose the symmetric gauge . But solving the
eqs. (1) and (2) give the following condition
\be \partial_\mu (B \sqrt {g})=0 \ee
As $B$ is constant , this equation shows that the symmetric gauge is possible
only when the surface is flat . Therefore $T_{\vec{\xi}}$ does not commute with
the Hamiltonian when the surface is not flat .

\section {Electron on sphere}

Consider an electron which is restricted to move on a sphere with radius R, in
the presence of the magnetic monopole at the center of the sphere. The flux of
the magnetic field $B$ is quantized by Dirac quantization condition
$B={{\hbar S} \over {eR^2}}$ ,
where S is half integer . The single-particle Hamiltonian is $^{[9]}$
\be H= {{{\bf{\Lambda}}^2}\over {2m R^2}}  \ \ ,\ee
where ${\bf{\Lambda}}= {\bf{r}} \times [-i h {\vec{\nabla}} + e{\bf{A}}]$ ,
${\bf{A}}$
satisfies ${\vec{\nabla}} \times {\bf{A}}= B{\bf{\Omega}}$ $({\bf{\Omega}}
={{\bf{R}}\over R}
)$ and ${\bf{\Lambda}}. {\bf{\Omega}}= 0$ . By using gauge freedom , the
electromagnetic potential can be taken
\be {\bf{A}}= -{{\hbar S}\over {eR}} ctg \theta \hat \varphi \ee
The eigenvalues of ${\bf{\Lambda}}^2$ are $(\ell(\ell +1)-S^2)\hbar
^2$ and the
first Landau Level is obtained for $\ell=s$ and is equal to ${{\hbar w_c}/ 2}$
$(w_c =eB/m)$ . It can be shown that the $\Lambda _i$'s satisfy
the following relations
\be [\Lambda_i , \Lambda_j]= i\hbar \epsilon _{ijk} (\Lambda_k - \hbar S
\Omega _ k) \ee
Let ${\bf{L}}= {\bf{\Lambda}} +\hbar s{\bf {\Omega}}$ , then
\be [ L_i, L_j]=i\hbar \epsilon _{ijk} L_k  \ \ .\ee
The operators $L_i$ are the generators of the symmetries of the Hamiltonian :
$[H, L_i]=0$ .

Now consider the group elements of this algebra
\be R_{\vec{\xi}}=e^{{i\over \hbar}{\bf{L}}.{{\vec{\xi}}\over R}} \hskip 1cm
and
\hskip 1cm R_{\vec{\eta}}= e^{{i\over\hbar}{\bf{L}}.{{\vec{\eta}}\over R}} \ee
where ${\vec{\xi}}=\xi \hat{\theta}$ and
${\vec{\eta}}=\eta\hat{\varphi}$ .
The product of
these operators is
\be R_{\vec{\xi}}R_{\vec{\eta}}=exp \{ {i\over{\hbar R}}({\vec{\xi}}
+{\vec{\eta}}).
{\bf{\Lambda}} -{1\over {2R^2\hbar
^2}}[{\vec{\xi}}.{\bf{\Lambda}},{\vec{\eta}}.
{\bf{\Lambda}}] +{i\over {3\hbar R}}[{\vec{\xi}}.{\bf{\Lambda}},M] + ...\} \
,\ee
where $M$ denotes the second term of the exponent . A simple calculation
shows that
$$[{\vec{\xi}}.{\bf{\Lambda}}, {\vec{\eta}}.{\bf{\Lambda}}]=-i\hbar \xi \eta
(\hbar
S +cotg \theta \Lambda _{\theta}) \ \ ,$$
where $\Lambda _\theta=\hbar (S ctg \theta + {i\over {sin \theta}} {{\partial}
\over {\partial \varphi}})$ . Now the equality ${\vec{\xi}}.{\bf{\Lambda}}=\xi
\Lambda_\theta$ implies that
 $[{\vec{\xi}}.{\bf{\Lambda}},M]=0$ . Therefore
$$ R_{{\vec{\xi}}}R_{{\vec{\eta}}}=exp \{ {i\over{\hbar R}}({\vec{\xi}} +
{\vec{\eta}}).
{\bf{\Lambda}} +{{i\xi \eta}\over {2R^2\sin ^2\theta}}(S+\cos \theta
{{\partial}
\over {\partial \varphi}})\} .$$
If we restrict ourselves to the region $\theta ={{\pi}\over 2}$ , we find that
\be R_{{\vec{\xi}}} R_{{\vec{\eta}}}= e^{{ieB\xi \eta}\over {2\hbar}}
R_{{\vec{\xi}}+{\vec{\eta}}} \ \ .\ee
In this way we recover the magnetic translation algebra and therefore the
$su_q(2)$ algebra with $q=exp ({{ie}\over {2\hbar}}{\bf{B}}.(\vec{\xi} \times
{\vec{\eta}}))$ .
This result shows that one can generate the $su_q(2)$ algebra by the
contraction
of the
$SU(2)$ group elements . This result has origin in the fact that the
contraction
of $su(2)$ algebra leads to Heisenberg algebra . Let $H, X^{\pm}$ be the
generators of $su(2)$ :
$$[H,X^{\pm}]=\pm 2X^{\pm}, \hskip 1.5cm [X^+, X^-]=H \ \ .$$
Assume $H^{\prime}$ and $P^\pm$ in $H=H^\prime +{1/{\epsilon ^2}}$
and $P^{\pm}=\epsilon X^{\pm}$ . Then
$$[H^{\prime}, P^{\pm}]= \pm 2P^{\pm} \hskip 0.5cm, \hskip 0.5cm [P^+, P^-]=
\epsilon^2 H+1 \ \ .$$
At the limit $\epsilon \rightarrow 0$ we have
$$[H^{\prime}, P^{\pm}]= \pm 2P^{\pm} \hskip 0.5cm, \hskip 0.5cm [P^+, P^-]=1
\ \ , $$
which is the Heisenberg algebra . Now the magnetic translation operator in the
plane has the following expression in terms of $ b=2p_{\bar z}-i{B\over 2}
z$ and $b^\dagger=2p_{ z}+i{B\over 2}\bar z$ $^{[4]}$
$$T_{{\bf{a}}}=W_{n,\bar{n}}=exp({1\over 2}(nb^\dagger - \bar{n} b))$$
where the commutation relation of the operators $b$ and $b^\dagger$ is
$[b,b^\dagger]=2B$. Therefore $T_{{\bf{a}}}$ are the exponential of the
Heisenberg
algebra .
But on the sphere $R_{{\vec{\xi}}}$ are the exponential of the
$su(2)$ algebra . So the above contraction of $su(2)$ to Heisenberg algebra
shed some light on
the reduction of the algebra of $R_{{\vec{\xi}}}$ to the $su_q(2)$ algebra .

\section {Electron on the Poincare upper half plane}

In this section we consider the Poincare upper half plane $H=\{ z=x+iy , y
>0\}$
, with the following metric
$$ds^2={dx^2+dy^2 \over y^2}.$$
For a covarient constant magnetic field $B$ , a paticular gauge choice leads
to
$$A_z=A_{\bar z}={B\over 2y}.$$
In this gauge the Hamiltonian (1) reduces to ( for simplicity we take $m=2$ )
\be H=-y^2\partial \bar{\partial}+{iB\over 2}y(\partial+\bar{\partial})
+{B^2\over 4} \ \ , \ee
and the ground states with energy $B/4$ are given by the solutions of the
equation
\be\nabla \psi_0=(\bar{\partial}+{B\over 2iy})\psi_0=0 \ \ ,\ee
which are $\psi_0(z,\bar {z})=y^B\psi_0(z)$ .

It can be easily checked that there are two operators which commutes with the
Hamiltonian (11) :
\be L_1=\partial +\bar{\partial}=\partial_x \ \ \ \ \ , \ \ \ \ \ L_2=z\partial
+
\bar{z}\bar{\partial} =x\partial_x+y\partial_y .\ee
$L_1$ and $L_2$ are the generators of a subalgebra of $sl(2,R)$ . Now let
$b=L_1$ and $b^\dagger =L_1^{-1}L_2 $ where $[b,b^\dagger ]=1$ . Then we can
choose the ground state wave functions to be the eigenfunctions of
$b^\dagger$ . By a direct calculation it can be shown that
$$b^\dagger \psi_0(\lambda \vert z,\bar{z}) =\lambda
\psi_0(\lambda \vert z,\bar{z}) \ \ , $$
where
\be \psi_0(\lambda \vert z,\bar{z})=y^B(\lambda -z)^{-B} \ \ .\ee
Then if we consider the symmetry operator $T_{\vec{\xi}}=e^{\xi_1b+\xi_2
b^\dagger}$ it can be shown that :
\be T_{\vec{\xi}}\psi_0(\lambda \vert z,\bar{z})=e^{\xi_2\lambda-{1\over
2}\xi_1
\xi_2}\psi_0(\lambda -\xi_1 \vert z,\bar{z}).\ee
Now it can be verified that the generators of $su_q(2)$ are
$$J_+={T_{{\vec{\xi}}}-T_{{\vec{\eta}}}\over q-q^{-1}} \  \ \ \ \ ,  \ \ \ \  \
\
J_-={T_{-{\vec{\xi}}}-T_{-{\vec{\eta}}}\over q-q^{-1}}$$
\be q^{2J_0}=T_{{\vec{\xi}}-{\vec{\eta}}} \ \ \ \ \ where  \ \ \ \ \
q=exp({1\over 2} {\vec{\xi}} \times {\vec{\eta}})  \ \ ,\ee
and the ground states wavefunctions are a representation of this algebra
\pagebreak
$$J_+\psi_0(\lambda \vert z,\bar{z})=[1/2-\lambda / \xi_1]_q
\psi_0(\lambda -\xi_1\vert z,\bar{z}) $$
$$J_-\psi_0(\lambda \vert z,\bar{z})=[1/2+\lambda / \xi_1]_q
\psi_0(\lambda +\xi_1\vert z,\bar{z}) $$
\be q^{\pm J_0}\psi_0(\lambda \vert z,\bar{z})
=q^{\mp\lambda /\xi_1}\psi_0(\lambda \vert z,\bar{z}),\ee
where the quantum symbol $[x]_q$ is defined by
$$[x]_q={q^x-q^{-x}\over q-q^{-1}}$$
In this way we showed the quantum group symmetry of QHE on the Poincare
upper half plane .
\\
{\bf {Acknowledgement}}
\\
We would like to thank professor H.Arfaei for valuable discussions.

\end{document}